\documentclass{amsart}
\usepackage{graphicx,cite,amsmath}
\newtheorem{thm}{Theorem}[section]
\def\beq{\begin{eqnarray}}
\def\eeq{\end{eqnarray}}
\def\bsp{\begin{split}}
\def\esp{\end{split}}

\def\d{\mathrm{d}}
\newcommand{\mf}[1]{{\mathfrak #1}}

\newcommand{\mb}[1]{{\mathbb #1}}
\newcommand{\mc}[1]{{\mathcal #1}}
\newcommand{\mbold}[1]{\mbox{\boldmath{\ensuremath{#1}}}}

\begin{document}

\title{\textbf{Solvegeometry gravitational waves}}
\author{Sigbj{\o}rn Hervik}%
\address{DAMTP, Cambridge University, Wilberforce Rd., Cambridge CB3 0WA, UK \textit{and} \linebreak Department of Mathematics \& Statistics, Dalhousie University, 
Halifax, Nova Scotia,
Canada B3H 3J5}%
\email{S.Hervik@damtp.cam.ac.uk,~ ~herviks@mathstat.dal.ca}%
\date{\today}
\begin{abstract}
In this paper we construct negatively curved Einstein spaces
describing gravitational waves having a solvegeometry wave-front
(i.e., the wave-fronts are solvable Lie groups equipped with a
left-invariant metric). Using the Einstein solvmanifolds (i.e.,
solvable Lie groups considered as manifolds) constructed in a previous paper as a starting point, we show that there also exist solvegeometry gravitational waves. Some  geometric aspects are discussed and examples of spacetimes having additional symmetries are given, for example, spacetimes generalising the Kaigorodov solution. The solvegeometry gravitational waves are also examples of spacetimes which are indistinguishable by considering the scalar curvature invariants alone.
\end{abstract}
\maketitle

\section{Introduction}
Recently we discussed how we could systematically construct higher-dimensional homogeneous Einstein solvmanifolds with negative curvature \cite{solvmanifolds}. These spaces generalise the Anti-de Sitter (AdS) spaces to non-maximally symmetric spaces but still possess much of their fascinating structure. As an example, we showed that these spaces allow for black hole solutions with a great variety of horizon geometries. In general negatively curved spaces are remarkably diverse \cite{thurston,thur:97,bp,goldman}; some of which is revealed in the rich structure of AdS spaces. Here, we shall show that more of this structure found in the AdS spaces survives into the more general Einstein solvmanifolds. 

For Riemannian spaces\footnote{In this paper Riemannian means that the metric has signature $(+++...+)$.} the Einstein solvmanifolds naturally enter the problem of classifying homogeneous Einstein manifolds\cite{Besse}. In fact, all known examples of (Riemannian) homogeneous Einstein manifolds with negative curvature are so-called \emph{standard} Einstein solvmanifolds \cite{Heber}; however, one does not know whether \emph{all} negatively curved homogeneous Einstein manifolds are standard solvmanifolds. Furthermore, a theorem by Pr\"ufer, Tricerri and Vanhecke \cite{PTV} states that two homogeneous Riemannian manifolds with identical scalar curvature invariants must be locally isometric. This is in contrast to the Lorentzian case.  We will see that the  Lorentzian case allows for two non-isometric homogeneous manifolds to have identical curvature invariants.

Almost two decades ago, Siklos found an interesting set of solutions
describing Lobatchevski-plane gravitational waves \cite{Siklos}. Here
we shall discuss the analogues of these solutions for the Einstein
solvmanifolds. The  solutions we find describe gravitational waves
where the wave-fronts are solvegeometries. Einstein spaces describing
gravitational waves have been fairly popular in the recent years. In
particular, the vacuum plane-wave spacetimes have some very
interesting physical and mathematical properties which have captured
the interest of a broad spectrum of researchers \cite{Guven, AmatiKlimcik, HS,
  Feinstein, GP, BFHP, Metsaev,BMN,HKL,EverExp}. 

These spaces may also be important in other higher-dimensional theories of gravity. For example, in the the brane-world scenario \cite{randall} our visible $(3+1)$-dimensional universe is a brane in a higher-dimensional bulk. This bulk is usually taken to be a negatively curved Einstein space; however, with a few exceptions, this bulk is usually taken to be a maximally symmetric AdS space. These solvmanifolds and their exact gravitational wave solutions may provide with further examples of possible bulks which can yield interesting behaviour of the brane dynamics \cite{branewaves}.  

In this paper we shall give a discussion of these solvegeometry gravitational waves. We  discuss some general properties of these solutions and provide  some  special cases which allow for additional symmetries. However,  first we will review some of the ideas behind the construction of homogeneous  Einstein solvmanifolds. 

\section{Review of Einstein Solvmanifolds}

Consider a real Lie algebra, $\mf{s}$, with the following properties:
 \begin{enumerate} 
\item{} The Iwasawa decomposition has the following orthogonal decomposition: 
\[ {\mf s}=\mf{a}\oplus\mf{n},\quad [\mf{s},\mf{s}]=\mf{n},\] 
where $\mf{a}$ is Abelian, and $\mf{n}$ is nilpotent.
\item{} All operators $\mathrm{ad}_{X}$, $X\in\mf{a}$ are symmetric.
\item{} For some $X^0\in\mf{a}$,
  $\left.\mathrm{ad}_{X^0}\right|_{\mf{n}}$ has positive eigenvalues.
\end{enumerate}
These Lie algebras
give rise to group manifolds, and since ${\mf{s}}$ is a solvable
  algebra, these are so-called
  \emph{solvmanifolds}. Using the left-invariant one-forms we can equip the groups with  left-invariant  metrics. Solvmanifolds with such left-invariant metrics will be called \emph{solvegeometries}. We will consider solvegeometries where the metric can  be written on the form
\beq 
\d s^2={ \d w}^2+\sum_{i=1}^ne^{-2q_iw}({\mbold\omega^i})^2,
\label{eq:genmetric}\eeq
where  ${\mbold\omega}^i$ are the left-invariant one-forms of a subalgebra of $\mf{s}$. 
These geometries are strong candidates for
Einstein spaces with negative curvature \cite{Wolter,Wolter2}. The simplest possible $\mf{s}$ \footnote{The simplest possible $\mf{s}$ is  the Lie algebra defined by $[{\bf X}^0,{\bf X}^i]={\bf X}^i, ~i=1...n$, with all other commutators being zero. The corresponding Lie group acts simply transitive on $(n+1)$-dimensional AdS space.} corresponds -- in the Lorentzian case -- to  AdS space.  thus general solvmanifolds with metrics (\ref{eq:genmetric}) can be considered as generalisations of the AdS spaces. 

The above form of the metric is particularly useful. The one-forms ${\mbold\omega^i}$ are themselves left-invariant one-forms on a group manifold. Hence, the geometry, $\mc{H}$, of the horospheres -- i.e. the hypersurfaces characterised by $w=$ constant -- can, up to a homothety, be given by the metric 
\beq
\d\Sigma^2=\sum_{i=1}^n({\mbold\omega^i})^2.
\eeq
It proves convenient to characterise the Einstein solvmanifolds by the geometry of the horospheres. We will write $M(\mc{H})$ for the solvegeometry with horospheres $\mc{H}$. We will also only consider Einstein spaces; i.e. 
\beq
R_{\mu\nu}=-\alpha^2 g_{\mu\nu}. 
\label{eq:alpha}\eeq
However, it should be noted that solvegeometries in general need not be Einstein. 

At this stage we should comment on the relation between the Riemannian and Lorentzian cases. The above construction is particularly useful for the Riemannian case where some precise uniqueness results are known \cite{Heber}. In general there is no obvious way of going from the Riemannian case to the Lorentzian case; however, if the horosphere geometry decomposes as $\mb{R}\times\mc{H}$, then we can perform a Wick rotation to a Lorentzian Einstein solvmanifold by $(i\mb{R})\times\mc{H}$. We will call such a Lorentzian Einstein solvmanifold which is obtained from a Riemannian one for \emph{regular}. 

In a previous paper we gave a recipe how to construct such spaces, and also gave some useful tools which simplify their construction considerably. An important result is that we could, for example, construct spaces of the type $M(\mb{E}^2\times \mc{H})$. It is these that will be of our concern here. These spaces have a regular Lorentzian Einstein analogue for which the metric can be written (introducing null coordinates $u,~v$):
\beq
\widetilde{\d s}^2=2e^{-2pw}\d v\d u+\d w^2+\sum_ie^{-2q_iw}\left({\mbold\omega}^i\right)^2.
\eeq
Note that 
\[\dim \mc{H}=\dim M( \mb{E}^2\times \mc{H})-3,\]
so the only 4-dimensional example is $\mc{H}=\mb{E}^1$. This is thus the well-known AdS$_4$. In dimension 5, we have $\dim\mc{H}=2$, so $\mc{H}$ can be characterised by the two distinct two-dimensional real Lie algebras. These give rise to the spaces AdS$_5$, and AdS$_3\times \mb{H}^2$ where $\mb{H}^2$ is the hyperbolic plane. In dimension 6 and higher, many non-trivial examples appear as the number of distinct real Lie algebras increases drastically. As an example we  provide with  a non-trivial 7-dimensional Einstein solvmanifold:
\beq
\d s^2&=&e^{-3qw}(2\d v\d u+\d X^2)+\d w^2+e^{-4qw}\left[\d x+\frac 12(y\d z-z\d y)\right]^2\nonumber \\ &&+e^{-2qw}(\d y^2+\d z^2),
\label{metricE3XNil}\eeq 
where $q=\sqrt{34}/17$. A more systematic approach to such Einstein solvmanifolds is given in the previous paper \cite{solvmanifolds}. 

We might wonder why the constant in eq.(\ref{eq:alpha}) is explicitly chosen to be negative. The reason is as follows. Any Riemannian left-invariant metric on a \emph{solvable} Lie group must have a non-positive Ricci curvature scalar; i.e. $R\leq 0$ \cite{Milnor}. Hence,  the positively curved ones are ruled out by construction. On the other hand, it should be noted that some of these solvmanifolds do have positively curved analogues (for example the symmetric ones \cite{Helgason}). However, these are not solvmanifolds and do not have a simple horospherical interpretation -- like eq.(\ref{eq:genmetric}) -- which makes it more difficult to deal with in a systematic way (especially their Lorentzian extension). 

\section{Gravitational waves with Solvegeometry wave-fronts}
Inspired by the existence of ``Lobatchevski plane waves'' found by Siklos \cite{Siklos} we will investigate similar gravitational wave solutions for the solvegeometries. 

Our main observation is the following:\footnote{We will use a notation where Greek indices run over the full spacetime manifold, and $x^a= (w, x^i)$.} \\
{\sl Assume that we have a regular Lorentzian Einstein solvmanifold given by the metric
\beq
\widetilde{\d s}^2=2e^{-2pw}\d v\d u+\d w^2+\sum_ie^{-2q_iw}\left({\mbold\omega}^i\right)^2, \quad \widetilde{R}_{\mu\nu}=-\alpha^2\widetilde{g}_{\mu\nu}, 
\label{eq:ESol}\eeq
and a function $H(u,x^a)$ which solves
\beq
\Box H=0, 
\eeq
where $\Box H=(1/\sqrt{-g})\partial_{\mu}\left(\sqrt{-g}g^{\mu\nu}\partial_{\nu}H\right)$. 
Then the metric 
\beq
{\d s}^2=2e^{-2pw}\left(\d v\d u+H(u,x^a)\d u^2\right)+\d w^2+\sum_ie^{-2q_iw}\left({\mbold\omega}^i\right)^2,
\label{eq:genSol}\eeq
is  Einstein with $R_{\mu\nu}=-\alpha^2g_{\mu\nu}$.}

Note that since $H_{,v}=0$ and $g^{uu}=0$ we have $\widetilde{\Box}H=\Box H$. Moreover, introducing  the function $H(u,x^a)$ will generally break the homogeneity of the spacetime. However, the spacetime always possesses one Killing vector, namely ${\mbold\xi}=\partial_v$. 

That the metric (\ref{eq:genSol}) is Einstein can easily be shown by simply calculate the Ricci tensor from the metric. The expressions for the Ricci tensor are given in the Appendix. Furthermore, for the Einstein solvmanifolds, 
\[ p\left[\ln \sqrt{-g}\right]_{,w}=-p\left(2p+\sum_iq_i\right)=-\left(2p^2+\sum_iq_i^2\right)=-\alpha^2, \]
because $p=\sum_iq_i^2/(\sum_iq_i)$ in order for it to be an Einstein space. From these expressions we see that provided the metric (\ref{eq:ESol}) is an Einstein space, then the metric (\ref{eq:genSol}) will be  Einstein  if $\Box H=0$. 

The similarity between the metric (\ref{eq:genSol}) and the metrics in the Kundt class \cite{Kundt} is evident. The Kundt class describes gravitational waves with some very interesting properties. We can make this similarity even more manifest. Consider the vector field ${\bf k}=\partial_v$. This is \emph{null}, and, in addition, obeys the following: 
\beq
k^{\mu}_{~;\mu}=k_{(\mu;\nu)}k^{\mu;\nu}=k_{[\mu;\nu]}k^{\mu;\nu}=0, \quad k_{\mu;\nu}k^{\nu}=0.
\eeq
Hence, ${\bf k}$ is geodesic, non-diverging, non-twisting, and shear-free (but it is not covariantly constant). To see this explicitly, we do a change of coordinates
\beq (V,U,X^a)&=&(e^{-2pw}v,u,x^a) \nonumber \\
\d s^2&=& 2\d U\left(\d V+2pV\d W+\bar{H}\d U\right)+g_{ab}\d X^a\d X^b, 
\label{eq:Kundtform}\eeq
where $\bar{H}\equiv e^{-2pw}H$. If we compare this with the metric (31.4) in \cite{Kramer} we note the similarity with the Kundt class explicitly. However, a notable difference is that for (\ref{eq:Kundtform}) Einstein, the metric $g_{ab}$ does not need to be of constant curvature (in dimension 4 it has to be); in general it is a solvegeometry which does not need to be even Einstein. In the 4-dimensional case it is known that the Siklos spacetimes are particular types of Kundt waves as shown by other authors \cite{ORR,BP,P}.
It is indeed remarkable that the similar construction goes through for the Einstein solvmanifolds. It is clear that solvmanifolds possess a remarkably nice geometric structure.   

Considering the wave-fronts of the gravitational waves, we note that these are also solvegeometries. By introducing the null frame 
\beq
{\mbold\ell}=\d u, \quad {\bf n}=e^{-2pw}(\d v+H\d u), \quad {\bf m}^a=\begin{cases} \d w, & a=w \\
e^{-q_iw}{\mbold\omega^i} & a=i,\end{cases}
\eeq
we see that the wave-fronts are given by the span of the ${\bf m}^a$'s. These wave-fronts are orthogonal to the ray congruence and form solvmanifolds which are not necessary Einstein. In fact, the wave-fronts are only Einstein if and only if the metric (\ref{eq:ESol}) is AdS. Moreover, the wave-fronts are totally geodesic and rigid with respect to an orthonormal vector field. The similarity to the Lobatchevski plane waves is thus clear and the metric (\ref{eq:genSol}) describes \emph{solvegeometry-plane gravitational waves}. 

\subsection{Electromagnetic fields} 
Both  the Lobatchevski plane waves and the ordinary vacuum plane waves  allow for a simple generalisation to include electromagnetic fields. The correspondence between Maxwell fields in Minkowski space and plane gravitational waves was noted already by Kundt \cite{Kundt}. A similar correspondence exists in this case.

Consider an electromagnetic one-form potential 
\beq
{\bf A}=\phi(u,x^a)\d u
\eeq
Then, choosing the Lorenz gauge $\d^{\dagger}{\bf A}=0$, the Maxwell equations  imply
\beq
\Box\phi=0.
\eeq
This is exactly the same as $H$ has to obey in the Einstein case. A solution to the Einstein-Maxwell equations can be found if we in addition require
\beq
\Box H=-e^{2pw}\phi^{,a}\phi_{,a}.
\eeq
This is completely analogous to the ordinary vacuum plane waves which has been extensively studied in the literature \cite{Kramer}. 

\subsection{The $\alpha\rightarrow 0$ limit}
Let us briefly discuss the dependence of the metric (\ref{eq:genSol}) on the parameter $\alpha$. Note first that by rescaling the metric (\ref{eq:genSol}) we can make the value of $\alpha$ anything we like. However, by writing the metric on the form  (\ref{eq:genSol}), we get from the Einstein equations
\beq
2p^2+\sum_iq_i^2=\alpha^2.
\eeq
Hence, in the limit $\alpha\rightarrow 0$, $p,q_i\rightarrow 0$. Furthermore, from the Einstein equations $\mathcal{R}_{ab}\rightarrow 0$, where  $\mathcal{R}_{ab}$ is the Ricci tensor of the wave-fronts. Thus we get that $\mathcal{R}_{ab}$ has to be Ricci-flat in the limit $\alpha\rightarrow 0$. In our construction, $\mathcal{R}_{ab}$ is the Ricci tensor of a \emph{homogeneous Riemannian space}. Hence, due to a result by Alekseevskii and Kimelfeld \cite{AK} which says that all Riemannian homogeneous Ricci-flat manifolds are locally flat, the $\alpha\rightarrow 0$ limit describes plane-fronted waves. Thus by making the function $H$ appropriately $\alpha$-independent we obtain the  well-known vacuum plane waves in the $\alpha\rightarrow 0$ limit. 
\footnote{In principle, there are nothing that prevents us from considering gravitational waves with non-homogeneous wave-fronts (some of these were discussed in \cite{solvmanifolds}). These would then have given more general vacuum Ricci-flat gravitational waves in the $\alpha\rightarrow 0$ limit.}

\subsection{Spacetimes with additional symmetries}
For general $H$ the metric (\ref{eq:genSol}) has only one Killing vector, ${\mbold\xi}=\partial_v$. However, if the function $H$ is of a particular form the metric can admit many more Killing vectors, and in some cases the metric (\ref{eq:genSol}) is even homogeneous. Here in this section we shall address the issue of Killing vectors and provide with some special cases which allow for additional symmetries. 

In a coordinate basis the Killing equations, $\xi_{(\mu;\nu)}=0$, reduce to 
\beq
\xi^u_{~,v} &=&0 \\
\xi^v_{~,v}+\xi^u_{~,u}-2p\xi^w &=&0\\
2H\xi^u_{~,u}+\xi^v_{~,u}-2pH\xi^w+H_{,a}\xi^a+H_{,u}\xi^u &=&0\\
2H\xi^u_{~,a}+\xi^v_{~,a}+e^{2pw}g_{ab}\xi^b_{~,u} &=& 0\\
\xi^u_{~,a}+e^{2pw}g_{ab}\xi^b_{~,v}&=& 0\\
g_{ac}\xi^c_{~,b}+g_{bc}\xi^c_{~,a}+g_{ab,c}\xi^c&=& 0.\label{eq:Killingab}
\eeq
Since the wave-fronts -- given by the metric $g_{ab}$ -- are themselves solvegeometries, there exists a set of vectors, ${\mbold\chi}_A$, acting transitively on the wave-fronts. These vectors will obey eq. (\ref{eq:Killingab}) but not necessary the remaining Killing equations. The form of these vectors, and the number of them, depends on the geometry of the wave-fronts. In many cases, these vectors are multiple transitive, for example, for the metric\footnote{By inspection we see that the five-dimensional space given by the metric (\ref{metricE3XNil}) with $u, v=$ constant has 6 Killing vectors.} (\ref{metricE3XNil}). In the following a generic vector of this type will be denoted by ${\mbold\chi}_A^{\epsilon}=\epsilon\partial_w+\chi^i_A\partial_i$ (in our case, $\epsilon=0$ or $1$). 

Depending on the form of $H(u,x^a)$, the following vectors can be Killing vectors,
\beq
{\mbold\xi}_1&=&\partial_v, \nonumber \\
{\mbold\xi}_2&=&\partial_u, \nonumber \\
{\mbold\xi}_3&=&u\partial_u-v\partial_v, \nonumber \\
{\mbold\xi}_4&=&p(1-\tilde{h})u\partial_u+p(1+\tilde{h})v\partial_v+{\mbold\chi}^1_A,\nonumber \\
{\mbold\xi}_5&=&pu^2\partial_u+\xi^v(x^a)\partial_v+u{\mbold\chi}^1_A\partial_i, \quad \xi^v_{~,a}+e^{2pw}g_{ab}\chi^b_A=0,\nonumber \\
{\mbold\xi}_6&=&-u\partial_u+v\partial_v+{\mbold\chi}^{0}_A, \nonumber \\
{\mbold\xi}_7&=&\xi^v(x^i)\partial_v+u{\mbold\chi}^0_A, \qquad \xi^v_{~,i}+e^{2pw}g_{ij}\chi^j_A=0,\nonumber \\
{\mbold\xi}_8 &=& {\mbold\chi}^{0}_A.
\eeq
Note that the vectors ${\mbold\xi}_5$ and ${\mbold\xi}_7$ do not always exist. A criterion for existence is that $\xi^v_{~,ab}=-(e^{2pw}g_{ac}\chi^c_A)_{,b}$ is symmetric in $a$ and $b$. 
\begin{table}[tbp]
\centering
\begin{tabular}{|c|c|c|}
\hline
$H$ & Einstein Condition & Killing Vectors \\ \hline \hline
$H(u,x^a)$ & $\Box H=0$ & ${\mbold\xi}_1$ \\ \hline
$A(x^a)$ & $\Box A=0$ & ${\mbold\xi}_1$, ${\mbold\xi}_2$  \\ \hline
$u^{-2}A(x^a)$ & $\Box A=0$ & ${\mbold\xi}_1$, ${\mbold\xi}_3$  \\ \hline
$\tilde{A}(x^a)$ & $\Box\tilde{A}=0$ & ${\mbold\xi}_1$, ${\mbold\xi}_2$, ${\mbold\xi}_6 $ \\ \hline
$A(u,w)$ & $A(u)e^{\sigma w}$ &  ${\mbold\xi}_1$, (${\mbold\xi}_7$), ${\mbold\xi}_8$ \\ \hline 
$A(w)$ & $e^{\sigma w}$ & ${\mbold\xi}_1$, ${\mbold\xi}_2$, (${\mbold\xi}_7$), ${\mbold\xi}_8$ \\ \hline 
$u^{-2(1+\beta)}F(e^{pw}u^{\beta})$ & $u^{(\Sigma_iq^2_i/p^2)\beta-2}e^{\sigma w}$ & ${\mbold\xi}_1$, ${\mbold\xi}_{4,\tilde{h}=1+1/\beta }$, (${\mbold\xi}_7$), ${\mbold\xi}_8$   \\ \hline 
$e^{2p\tilde{h}w}$ & $e^{\sigma w}$ & ${\mbold\xi}_1$, ${\mbold\xi}_2$, ${\mbold\xi}_{4,\tilde{h}=\sigma/2p}$, (${\mbold\xi}_7$),   ${\mbold\xi}_8$ \\ \hline
$ e^{-2pw} $ & $-$ & ${\mbold\xi}_1$, ${\mbold\xi}_2$, ${\mbold\xi}_{4, \tilde{h}=-1}$, (${\mbold\xi}_5$), (${\mbold\xi}_7$), ${\mbold\xi}_8$ \\ \hline
\end{tabular}
\caption{Table with particular forms of $H(u,x^a)$ and their Killing vectors. Here, $\sigma=2p+\sum_iq_i$, and a vector in brackets means the vector is a Killing vector if it exists. The function $\tilde{A}(x^a)$ is a function obeying $2\tilde{A}+\chi_A^a\tilde{A}_{,a}=0$.}
\label{KillingVectors}
\end{table}

A sample of special cases for which the metric allows for additional Killing vectors is given in Table \ref{KillingVectors}. This is not an exhaustive list of all the possible cases since this would require a knowledge of all the possible Einstein solvmanifolds and their geometric structure. In \cite{Siklos} the Lobatchevski plane waves  and in \cite{SGP} the Kundt waves are considered in greater detail. 

There are some cases worth mentioning. The case $H=e^{\sigma w}$  generalises the known Kaigorodov spaces \cite{Kaigorodov}. These allow for a transitive group of isometries and are thus homogeneous. Note that the three vectors ${\mbold\xi}_1$, ${\mbold\xi}_2$, ${\mbold\xi}_4$ span a three-dimensional Lie algebra isomorphic to the Bianchi type VI$_h$ Lie algebra \cite{bianchi,EM} (where  $h=-1/\tilde{h}^2$). In addition to these, there is a  set of Killing vectors acting transitively on the horospheres of the wave-front, given by different ${\mbold\xi}_8$. These commute with  ${\mbold\xi}_1$ and  ${\mbold\xi}_2$, but does not commute with ${\mbold\xi}_4$. The set of simply transitive Killing vectors forms a solvable algebra; hence, the generalised  Kaigorodov spaces are solvegeometries.

Another interesting case is when $H=e^{-2pw}$. These spaces generalise the Defrise solution \cite{Defrise}. Note that they are not Einstein spaces  since $\Box H\neq 0$. Notwithstanding this, they are homogeneous.

\subsection{The curvature invariants}  Another interesting fact about these spacetimes is a remarkable property regarding  their scalar curvature invariants. All the scalar curvature invariants of the metric (\ref{eq:genSol}) are not only constants, but they are also completely independent of the function $H$. Hence, two spacetimes of the same form (\ref{eq:genSol}) but with different functions $H$ are indistinguishable by considering the scalar curvature invariants alone \cite{HP}. This is an astonishing fact which may have interesting consequences for their physical properties. A similar phenomenon happens for the vacuum plane-wave spacetimes for which the scalar curvature invariants all vanish \cite{PPCM}. 

These results are in stark contrast to the Riemannian case. For Riemannian manifolds we have the following theorem:
\begin{thm}[Pr\"ufer-Tricerri-Vanhecke \cite{PTV}]
Let $(M,g)$ be a Riemannian manifold such that all scalar curvature invariants are constants. Then $(M,g)$ is locally homogeneous. Moreover, $(M,g)$ is uniquely determined by these invariants up to a local isometry.
\end{thm} 
The gravitational wave spacetimes are therefore counterexamples to the above theorem for Lorentzian manifolds. In fact, uniqueness even fails in the Lorentzian case after assuming $(M,g)$ is homogeneous \emph{and} Einstein; the regular Lorentzian Einstein manifold and the corresponding Kaigorodov spacetime are both homogeneous, Einstein, and have identical scalar curvature invariants. 

\section{Discussion}
In this paper we have considered a class of negatively curved Einstein manifolds describing gravitational waves. The motivation of their existence  came from the the known Lobatchevski plane waves \cite{Siklos} which are the simplest non-trivial solutions of this type. The solutions described here can be interpreted as \emph{solvegeometry-plane gravitational waves}. 

We  discussed some of the geometrical properties of these solutions. In particular, some special cases admitting additional symmetries were investigated. Another interesting fact about these spacetimes is the remarkable property regarding  their scalar curvature invariants. As we pointed out, all the scalar curvature invariants of the metric (\ref{eq:genSol}) are not only constants, but they are also completely independent of the function $H$. Hence, two spacetimes of the same form (\ref{eq:genSol}) but with different functions $H$ are indistinguishable by considering the scalar curvature invariants alone. We might contemplate what consequences this have for, for example, higher curvature  theories of gravity. These theories  are usually based on an action principle involving scalar curvature invariants. Hence, since these spaces all have identical curvature invariants they might yield non-trivial solutions to some higher-curvature theories of gravity.  

This work has given more evidence for the incredible richness and diversity  of negatively curved spaces in general and of Einstein solvmanifolds in particular. It seems like we so far only understand a fraction of what these solvmanifolds have to offer and that many more hidden secrets await to be discovered. 

\section*{Acknowledgments}
This work was funded by the Norwegian Research Council, an Isaac Newton Studentship and an AARMS PostDoctoral Fellowship. 

\appendix
\section{Curvature properties}
Assume that we have the metric 
\beq
{\d s}^2=2e^{-2pw}\left(\d v\d u+H(u,x^a)\d u^2\right)+\d w^2+\sum_ie^{-2q_iw}\left({\mbold\omega}^i\right)^2,
\eeq
which for $H=0$ reduces to a homogeneous solvmanifold in horospherical coordinates. In the following we will introduce a \emph{coordinate basis}, given by the coordinates $u,v,x^a$. Quantities with tildes refer to the case $H=0$. 
Using the formula for the Christoffel symbols, 
\beq
\Gamma^{\lambda}_{~\alpha\beta}=\frac 12g^{\lambda\mu}\left(g_{\mu\alpha,\beta}+g_{\mu\beta,\alpha}-g_{\alpha\beta,\mu}\right),
\eeq
we find the non-zero Christoffel symbols (symmetric in the lower indices):
\beq
&&\Gamma^v_{~uu}=H_{,u}, \quad \Gamma^v_{~ua}=H_{,a} \\
&&\Gamma^v_{~vw}=-p, \quad \Gamma^u_{~uw}=-p,\quad \Gamma^w_{~uv}=pe^{-2pw},\\ 
&&\Gamma^a_{~uu}=-e^{-2pw}\left(-2pH\delta^a_{~w}+g^{ab}H_{,b}\right),\quad \Gamma^a_{~bc}=\widetilde{\Gamma}^a_{~bc}.
\eeq
The Ricci tensor is given by the Christoffel symbols as
\beq
R_{\alpha\beta}=\Gamma^{\mu}_{~\alpha\beta,\mu}-\Gamma^{\mu}_{~\alpha\mu,\beta}
+\Gamma^{\mu}_{~\alpha\beta}\Gamma^{\rho}_{~\mu\rho}-\Gamma^{\rho}_{~\alpha\mu}\Gamma^{\mu}_{~\rho\beta},
\eeq
which gives
\beq
R_{uu}&=&g_{uu}p\left[\ln\sqrt{-g}\right]_{,w}-e^{-2pw}\frac{1}{\sqrt{-g}}\left(\sqrt{-g}g^{ab}H_{,a}\right)_{,b}\nonumber \\
R_{uv}&=& g_{uv}p\left[\ln\sqrt{-g}\right]_{,w} \nonumber \\
R_{vv}&=& 0 \nonumber \\
R_{ab}&=& \widetilde{R}_{ab}.
\eeq
Note that since $H_{,v}=0$, we have 
\beq
\frac{1}{\sqrt{-g}}\left(\sqrt{-g}g^{ab}H_{,a}\right)_{,b}=\frac{1}{\sqrt{-g}}\left(\sqrt{-g}g^{\mu\nu}H_{,\mu}\right)_{,\nu}\equiv\Box H.
\eeq 

\end{document}